\documentclass[reprint,amsmath,amssymb]{revtex4-2}

\usepackage{tikz}
\usepackage{pgfplots}
\usepackage{circuitikz}

\usepackage{hyperref}
\usepackage{graphicx,amsmath,amsfonts}
\usepackage{amssymb}
\usepackage{dcolumn}
\usepackage{mathrsfs}
\usepackage{xcolor}
\usepackage{bm}
\usepackage{enumerate}
\usepackage{pgfplots}
\usepackage{tikz}
\usetikzlibrary{decorations.markings}
\usetikzlibrary{plotmarks}
\usepackage[export]{adjustbox}

\renewcommand{\vec}[1]{\mathbf{#1}}
\newcommand{\vers}[1]{\hat{\mathbf{#1}}}
\newcommand{\dyad}[1]{\overset{\text{\scalebox{.7}{$\boldsymbol{\leftrightarrow}$}}}{\mathbf{#1}} }
\newcommand{\dS}{\, \mathrm{d}S}
\newcommand{\Npart}{p}
\newcommand{\partp}{\mu}
\newcommand{\partq}{\nu}

\newlength{\mywidth}
\newlength{\myheight}
\newlength{\mydelta}
\begin{document}

\title{Multilevel Fast Multipole Algorithm \\ for Electromagnetic Scattering by Large Metasurfaces \\ using Static Mode Representation}

\author{Emanuele Corsaro}
\affiliation{Department of Electrical Engineering and Information Technology, Universit\`{a} degli Studi di Napoli Federico II, via Claudio 21,  Napoli, 80125, Italy}
 
\author{Giovanni Miano}
\affiliation{Department of Electrical Engineering and Information Technology, Universit\`{a} degli Studi di Napoli Federico II, via Claudio 21,  Napoli, 80125, Italy}

\author{Antonello Tamburrino}
\affiliation{Università degli Studi di Cassino e del Lazio Meridionale, Via G. Di Biasio n. 43, 03043 Cassino (FR), Italy}

\author{Salvatore Ventre}
\affiliation{Università degli Studi di Cassino e del Lazio Meridionale, Via G. Di Biasio n. 43, 03043 Cassino (FR), Italy}
 
\author{Carlo Forestiere}
\email[]{carlo.forestiere@unina.it}
\affiliation{Department of Electrical Engineering and Information Technology, Universit\`{a} degli Studi di Napoli Federico II, via Claudio 21,  Napoli, 80125, Italy}

\begin{abstract}
Metasurfaces, consisting of large arrays of interacting subwavelength scatterers, pose significant challenges for general-purpose computational methods due to their large electric dimensions and multiscale nature. This paper introduces an efficient boundary element method specifically tailored for metasurfaces, leveraging the Poggio-Miller-Chang-Harrington-Wu-Tsai (PMCHWT) formulation. Our method combines the Multilevel Fast Multipole Algorithm (MLFMA) with a representation of the unknown equivalent surface current density by means of static modes, a set of entire domain basis functions dependent only on object shape but independent of the material and frequency. The compression of the number of unknowns enabled by the Static Mode Representation (SMR), combined with the \(\mathcal{O}(N \log N)\) complexity of MLFMA matrix-vector products, significantly reduces CPU time and memory requirements compared to classical MLFMA with RWG basis functions. We demonstrate the accuracy, time, and memory requirements of this method through several test cases including the full-wave simulation of a $100 \lambda \times 100 \lambda$ canonical metalens. The MLFMA-SMR method offers substantial benefits for the analysis and optimization of metasurfaces and metalenses.
\end{abstract}

\maketitle

\section{Introduction}

Metasurfaces are collections of interacting subwavelength scatterers, typically operating near resonance \cite{holloway_overview_2012}. By carefully designing the geometry and arrangement of these scatterers, metasurfaces can engineer a wide class of effective constitutive relations \cite{pfeiffer_metamaterial_2013}.  They are also versatile in shaping and manipulating electromagnetic waves for various applications, including analog computing \cite{silva_performing_2014}, biosensors \cite{altug_advances_2022}, and imaging \cite{chen_review_2016}. Metalenses are one common class of metasurfaces, in which the unit cells are designed to impart a specific, position-dependent phase shift to transmitted light, resulting in a global focusing effect  \cite{yu_flat_2014}. The accurate simulation of metasurfaces and metalenses, with efficient use of time and memory, is crucial for both understanding their fundamental properties \cite{choi_realization_2024}  and optimizing their performance \cite{kang_large-scale_2024}.

Metasurfaces design often relies on the ``unit cell” approximation \cite{yu_flat_2014,aieta_multiwavelength_2015,lalanne_prehistory_2023}. This approach constructs a library of the electromagnetic response of meta-atoms, such as disks, pillars, or nanofins, as a function of a design parameter, like the disk radius, pillar height, or the nanofin orientation, assuming the structure to be locally periodic. Then, the contributions of the different “unit-cell” are combined by linear superposition. Unfortunately, the “unit-cell” approach has several limitations: the most critical limitation is the assumption that the interaction between adjacent meta-atoms is assumed equal to that in periodic arrays, which necessitates a slowly changing structure. This constraint limits the design space, resulting in narrow-bandwidth operation, low efficiencies, and many types of non-ideality.

General-purpose electromagnetic simulation tools, based on either \textit{integral} or \textit{differential} formulations, face significant challenges in the simulation of metasurfaces \cite{choi_realization_2024}. Differential formulations, including finite-element methods and finite-difference methods, are easy to implement and result in sparse matrices but require truncation of the computational domain through the introduction of convenient boundary conditions \cite{jin_theory_2011}.   Among them, the finite-difference time-domain (FDTD) method \cite{yee_finite-difference_1997,taflove_computational_2005,oskooi_meep_2010}  is widely used to model metasurfaces. However, the finite resolution leads to inaccuracies in wave speed, resulting in phase accumulation errors as the size of the scattering region increases \cite{xue_fullwave_2023}. Hence, even though the complexity of FDTD simulations nominally increases linearly with the size of the array, maintaining a fixed error requires increasing the resolution \cite{shlager_selective_1995,taflove_computational_2005}. This makes the approach prohibitive when applied using traditional computing platforms. 

Integral formulations are appealing because they define the unknowns only within the objects' volumes or on their boundaries if the objects are spatially piecewise homogeneous, and naturally satisfy radiation conditions at infinity. However, they usually involve dense matrices, which are computationally demanding to invert, even with acceleration techniques like the fast multipole method \cite{greengard_fast_1987,chew_fast_2000,michielssen_multilevel_1996}, hierarchical matrices \cite{hackbusch_introductory_2015,hackbusch_h2-matrices_2000}, or QR decomposition \cite{poirier_numerically_1998}.

A brute-force strategy for scaling full-wave simulations involves adapting general-purpose computational electromagnetic algorithms to massive parallelization for CPU-based \cite{fostier_asynchronous_2008,donepudi_higher_2001,velamparambil_analysis_2005,ergul_fast_2013,waltz_massively_2007, fijany_massively_1995} or GPU-based computing platforms \cite{hughes_perspective_2021,he_solving_2022}.   However, there remains significant potential to refine these methods by leveraging the peculiar characteristics of electromagnetic scattering in metasurfaces, which generally involve repeated particle shapes, typically of subwavelength dimensions, arranged in aperiodic positions and arbitrary orientations. In this work, we pursue this strategy by proposing a fast boundary integral method specifically tailored to this problem. We combine the Multilevel Fast Multiple Method with the Poggio-Miller-Chang-Harrington-Wu-Tsai (PMCHWT)  \cite{chang_surface_1977, wu_scattering_1977, poggio_chapter_1973,harrington_field_1993} boundary integral equation  employing an expansion of the unknowns in terms of a particular class of entire-domain basis functions, defined on the individual meta-atom, and denoted as static modes. As demonstrated in Ref. \cite{forestiere_static_2023}, static modes drastically reduce the number of unknowns in scattering problems involving arrays of penetrable particles compared to discretization based on sub-domain basis functions, such as RWG.

This manuscript is organized as follows: In Section \ref{sec:PMCHWT}, we recall the PMCHWT formulation and its discretization. In Section \ref{sec:SMR}, we briefly review the static mode basis \cite{forestiere_static_2023}, and in Section \ref{sec:MLFMA}, we apply the fast multipole method to accelerate the solution of the PMCHWT where the unknown equivalent surface current densities are expanded in terms of static modes. In Section \ref{sec:Results}, we first validate the proposed method by comparing, for arrays of spheres, the results with those obtained using the multiparticle Mie theory \cite{xu_electromagnetic_1995}. We then assess the computational burden in terms of CPU time and memory requirements by comparing the static mode representation to a more traditional approach using RWG basis functions. Finally, we present results for three large particle arrays: a golden angle spiral array, a two-layer Moiré superlattice, and a $100 \lambda \times 100 \lambda$ canonical metalens.
\section{Surface Integral Equation}
 \label{sec:PMCHWT}
	Let us consider a system composed of $\Npart$ objects illuminated by a time-harmonic electromagnetic
	field $\mathrm{Re}\{\vec{E}_{\text{inc}} e^{j\omega t}\}$, $\mathrm{Re}\{\vec{H}_{\text{inc}} e^{j\omega t}\}$. Each object is made of a linear, homogeneous, isotropic material that occupies the three-dimensional domain $\Omega_\partp$, with $\partp = 1 ,..., \Npart$. This material is described by a permittivity
	$\varepsilon^+(\omega)$ and a permeability $\mu^+(\omega)$. The objects are surrounded by a background
	medium with permittivity $\varepsilon^-(\omega)$ and permeability
	$\mu^-(\omega)$. The equivalent electric $\vec{j}_e^\partp$
	and magnetic $\vec{j}_m^\partp$ surface current densities, defined on the boundary  $\partial \Omega_\partp$ of the domain $\Omega_\partp$, are
	solutions of the Poggio-Miller-Chang-Harrington-Wu-Tsai (PMCHWT) \cite{chang_surface_1977,wu_scattering_1977,poggio_chapter_1973,harrington_field_1993} surface integral problem

 \begin{equation}
	\begin{bmatrix}
		\mathcal Z^{11} &\cdots &\mathcal Z^{1\Npart} \\
		\vdots &\ddots &\vdots\\
		\mathcal Z^{\Npart 1} &\cdots &\mathcal Z^{\Npart \Npart}
	\end{bmatrix} \begin{bmatrix}
	\vec{j}^1 \\
	\vdots\\
	\vec{j}^{\Npart}
\end{bmatrix} = \begin{bmatrix}
\vec{v}^1 \\
\vdots\\
\vec{v}^{\Npart}
\end{bmatrix} ,
\label{eq:PMCHWT_1}
\end{equation}
	
with
\begin{equation}
	\mathcal{Z}^{\partp \partq} = \begin{bmatrix}
		\zeta^- \mathcal{T}_-^{\partp \partq} + \zeta^+ \mathcal{T}_+^{\partp \partq} \delta_{\partp \partq}  &\mathcal{K}_-^{\partp \partq} + \mathcal{K}_+^{\partp \partq}\delta_{\partp \partq}\\
		-(\mathcal{K}_-^{\partp \partq} + \mathcal{K}_+^{\partp \partq}\delta_{\partp \partq}^{\partp \partq}) & \frac{1}{\zeta^-} \mathcal{T}_-^{\partp \partq} + \frac{1}{\zeta^+} \mathcal{T}_+^{\partp \partq}\delta_{\partp \partq}
	\end{bmatrix},
\label{eq:PMCHWT}
\end{equation}
where $\delta_{\partp\partq}$ is the Kronecker delta
\begin{equation}
	\vec{j}^\partp = \begin{bmatrix}
		\vec{j}_e^\partp\\ \vec{j}_m^\partp
	\end{bmatrix}, \hspace{5ex}	\vec{v}^\partp = \begin{bmatrix}
	\vec{e}_0^\partp\\ \vec{h}_0^\partp
\end{bmatrix},
\end{equation}
	\begin{equation}
\vec{e}_0^\partp = - \vers{n}_\partp\times \vers{n}_\partp \times \vec{E}_{\text{inc}} |_{\partial \Omega_\partp}, \hspace{4ex} \vec{h}_0^\partp = - \vers{n}_\partp\times \vers{n}_\partp \times \vec{H}_{\text{inc}} |_{\partial \Omega_\partp},
\end{equation}
and $\hat{n}_\mu$ is the normal to the surface $\partial \Omega_\mu$, pointing outward. The operators $\mathcal{T}_\pm^{\partp \partq}$ and $\mathcal{K}_\pm^{\partp \partq}$ are, respectively, the EFIE and MFIE integral operators, defined as
\begin{align}
	&\mathcal{T}_\pm^{\partp \partq} \{\vec{u}\} (\vec{r}) =\nonumber \\
 &\frac{1}{jk^\pm} \vers{n}_\partp \times \vers{n}_\partp \times \int_{\partial \Omega_\partq} \nabla g^\pm (\vec{r} - \vec{r}') \nabla'_S \cdot \vec{u}(\vec{r}') \dS' +  \nonumber \\
  &j k^\pm \vers{n}_\partp\times \vers{n}_\partp \times \int_{\partial \Omega_\partq} g^\pm (\vec{r}-\vec{r}') \vec u(\vec r') \dS',
 \label{eq:EFIE_def}
\end{align}
\begin{align}
	\mathcal{K}_\pm^{\partp \partq} \{\vec{u}\}(\vec{r}) = \vers{n}_\partp \times \vers{n}_\partp \times \int_{\partial \Omega_\partq} \vec{u}(\vec{r}') \times \nabla' g^{\pm}(\vec{r} - \vec{r}') \dS',
 \label{eq:MFIE_def}
\end{align}
where $\nabla_S' \cdot $ denotes the surface divergence and $g^\pm$ is the homogeneous space Green's function of the region $\Omega_\pm$, i.e.
\begin{equation}
	g^\pm (\vec{r} - \vec{r}') = \frac{e^{-jk^\pm \|\vec r - \vec r'\|}}{4\pi \|\vec{r} - \vec{r}'\|},
\end{equation} 
with $k^\pm = \omega \sqrt{\mu^\pm \varepsilon^\pm}$ and $\zeta^\pm = \sqrt{\mu^\pm /\varepsilon^\pm}$.
 Apart from a multiplicative factor, the operator \(\mathcal{T}_\pm^{\partp \partq} \{\vec{u}\} (\vec{r})\) takes a surface electric current density \(\vec{u}\) defined on the boundary $\partial \Omega_\partq$ of the \(\nu\)-th particle and outputs the tangential component of the electric field it generates on the surface of the \(\mu\)-th particle. Similarly, the operator \(\mathcal{K}_\pm^{\partp \partq} \{\vec{u}\} (\vec{r})\) takes a surface magnetic current density \(\vec{u}\) defined on $\partial \Omega_\partq$ and outputs the tangential component of the electric field it generates on the surface of the \(\mu\)-th particle.

\subsection{Galerkin equations}
Aiming at the solution of the PMCWHT of Eq. \eqref{eq:PMCHWT_1}, we  represent the equivalent electric { $\left\{ \vec{j}_e^\partp \right\}_{\partp=1}^\Npart$ and magnetic $\left\{ \vec{j}_m^\partp \right\}_{\partp = 1}^{\Npart}$} surface current densities on the boundary $\partial \Omega_\partp$ of the $\mu$-th particle
as a linear combination of   basis functions $\mathcal{B}_\varphi^\partp = \{\boldsymbol{\varphi}^\partp_n(\vec{r})\}_{n = 1}^{N_\varphi}$:
\begin{equation}
\begin{cases}
	\displaystyle              
	\vec{j}_e^\partp(\vec r) = \sum_{n = 1}^{N_\varphi} \alpha^{\partp,e}_n \boldsymbol{\varphi}^\partp_n (\vec{r}) \\
	\displaystyle
	\vec{j}_m^\partp(\vec r) = \sum_{n = 1}^{N_\varphi} \alpha^{\partp,m}_n \boldsymbol{\varphi}_n^\partp (\vec{r}).
\end{cases}
\label{eq:galerkin}
\end{equation}

We find the finite-dimensional approximation of the PMCHWT by substituting (\ref{eq:galerkin}) in (\ref{eq:PMCHWT_1}) and by projecting along the same set of basis functions, according to the Galerkin projection scheme:
\begin{equation}
	\begin{bmatrix}
		Z^{11} &\cdots &Z^{1\Npart} \\
		\vdots &\ddots &\vdots\\
		Z^{\Npart 1} &\cdots &Z^{\Npart \Npart}
	\end{bmatrix} \begin{bmatrix}
	\vec{J}^1 \\
	\vdots\\
	\vec{J}^{\Npart}
\end{bmatrix} = \begin{bmatrix}
\vec{V}^1 \\
\vdots\\
\vec{V}^{\Npart}
\end{bmatrix} ,
\label{eq:PMCHWT_finite}
\end{equation}
where
\begin{equation}
	Z^{\partp\partq} = \begin{bmatrix}
		\zeta^- T_-^{\partp\partq} + \zeta^+ T_+^{\partp\partq} \delta_{\partp\partq}  &K_-^{\partp\partq} + K_+^{\partp\partq}\delta_{\partp\partq}\\
		-(K_-^{\partp\partq} + K_+^{\partp\partq}\delta_{\partp\partq}) & \frac{1}{\zeta^-} T_-^{\partp\partq} + \frac{1}{\zeta^+} T_+^{\partp\partq}\delta_{\partp\partq}
	\end{bmatrix},
 \label{eq:Zpq_delta}
\end{equation}

\begin{equation}
	[T_\pm^{\partp\partq}]_{mn} = \langle \boldsymbol{\varphi}_m^\partp | \mathcal{T}_\pm^{\partp\partq} | \boldsymbol{\varphi}_n^\partq \rangle, \hspace{5ex} 	[K_\pm^{\partp\partq}]_{mn} = \langle \boldsymbol{\varphi}_m^\partp | \mathcal{K}_\pm^{\partp\partq} | \boldsymbol{\varphi}_n^\partq \rangle,
 \label{eq:PMCHWT_Finite1}
\end{equation}

\begin{equation}
	\vec{J}^\partp= \begin{bmatrix}
		\vec J_e^\partp \\ \vec J_m^\partp
	\end{bmatrix}, \hspace{5ex}  \vec V^\partp = \begin{bmatrix}
	\vec{E}_0^\partp \\ \vec{H}_0^\partp
\end{bmatrix},
\end{equation}
\begin{equation}
	\vec{J}^\partp_{e} = [\alpha_1^{\partp,e}, \ldots ,\alpha_N^{\partp,e}]^{\intercal}, \hspace{5ex} 	\vec{J}_{m}^\partp = [\alpha_1^{\partp,m}, \ldots ,\alpha_N^{\partp,m}]^{\intercal},
\end{equation}

\begin{equation}
[\vec{E}_0^\partp]_n = \langle \boldsymbol{\varphi}_n^\partp | \vec{e}_0^\partp \rangle, \hspace{5ex} [\vec{H}_0^\partp]_n = \langle \boldsymbol{\varphi}^\partp_n | \vec{h}_0^\partp \rangle,
\end{equation}
\begin{equation}
	\langle \vec{u} | \vec{v} \rangle = \int_{\partial \Omega}  \vec{u}^*(\vec{r}) \cdot  \vec{v}(\vec{r}) \dS,
\end{equation}
If the particles have the same shape and are discretized in the same way, each block $Z^{\partp\partq}$ is a square matrix of dimension $2N_\varphi$. In this case, the finite-dimensional system (\ref{eq:PMCHWT_finite}) has $2N_\varphi \Npart$ degrees of freedom. 

We can choose as basis either \textit{sub-domain} or \textit{entire-domain} basis functions.
Sub-domain basis functions are non-zero only over a localized region of each individual object. A typical choice for these functions is the Rao-Wilton-Glisson (RWG) functions \cite{rao_electromagnetic_1982}. We denote the set of  sub-domain of  basis functions as $\mathcal{B}_\xi^\partp = \{ \boldsymbol{\xi}^\partp_n(\vec{r})\}_{n=1}^{N_\xi}$.  Instead, entire domain basis functions extend over the entire domain of each individual object.  These basis functions can be generated through methods like separation of variables (e.g., vector spherical or spheroidal basis functions) or by solving auxiliary eigenvalue problems, as in the case of characteristic modes \cite{garbacz_modal_1965,chang_surface_1977,harrington_characteristic_1972,chen_characteristic_2015}, (see \cite{bucci_use_1995,angiulli_characteristic_1998}. For further examples, see \cite{bucci_use_1995,angiulli_characteristic_1998} for arrays of perfectly conducting particles, \cite{faenzi_metasurface_2019} for perfectly conducting metasurfaces, and material-independent modes \cite{forestiere_material-independent_2016,pascale_full-wave_2019}.  We denote the set of entire-domain basis functions $\mathcal{B}_\psi^\partp = \{ \boldsymbol{\psi}^\partp_n(\vec{r})\}_{n=1}^{N_\psi}$.

When we use sub-domain basis functions the numerical solution quickly becomes prohibitive as the number of particles $\Npart$ increases, even with the implementation of acceleration techniques such as fast multipole algorithms. This is due to both computational and memory burdens.  To address this challenge, we turn to entire domain basis functions. These functions are related to the basis $\mathcal{B}_\xi^\partp$ through the linear transformation
\begin{equation}
	\boldsymbol{\psi}_i^\partp(\vec{r}) = \sum_{j=1}^{N_\xi} m_{ij} \boldsymbol{\xi}^\partp_j(\vec{r}),
\end{equation}
where $M = [m_{ij}]$ is a real-valued transformation matrix. To find the representations of the operators $T_{\pm}^{\partp\partq}$ and $K_{\pm}^{\partp\partq}$ in the new basis $\mathcal{B}_\psi$, we can transform them from the basis $\mathcal{B}_\xi$ representation using
\begin{equation}
	\{A_{\pm}^{\partp\partq}\}_\psi = M \{A_{\pm}^{\partp\partq}\}_\xi M^{\intercal},
	\label{eq:BasisTransformation}
\end{equation}
with $A \in \{ T, K \}$ .
 
\section{Static Current Density Modes as Entire domain basis }
\label{sec:SMR}

Guided by the fact that the particles constituting the metasurface, i.e., the meta-atoms, are typically smaller than the wavelength, we seek a set of entire-domain basis functions of an individual meta-atom that can efficiently describe the unknown { equivalent} electric and magnetic surface current densities in the small particle regime. Mathematically speaking, we aim for basis functions that diagonalize some of the operators involved in the PMCHWT formulation in the static limit. We also require that set be complete and thus capable of describing any square-integrable vector function defined on the surface of the object.  

A set of entire-domain basis functions, defined on the surface of an \textit{individual} meta-atom and termed \textit{static surface current density modes}, has recently been introduced and proven effective in representing surface current density fields  \cite{forestiere_electromagnetic_2019,forestiere_static_2023}. These modes consist of two distinct subsets: \textit{longitudinal} modes, characterized by a zero surface curl, and \textit{transverse} modes, with zero surface divergence. These subsets are constructed by introducing two separate auxiliary eigenvalue problems. These two eigenvalue problems are associated with two operators, $\mathcal{T}_0^\perp$ and $\mathcal{T}_0^\parallel$, which are obtained from the static-limit of the first and second operators in the expression of $\mathcal{T}$ \eqref{eq:EFIE_def}, respectively.

Let $\Omega$ be one of the meta-atom $\{\Omega_\partp\}_{\partp = 1}^\Npart$ in the array, whose boundary $\partial \Omega$ is ``sufficiently regular" \cite{monk_finite_2003}; $\vers{n}$ is the normal to $\partial \Omega$ pointing outward. The longitudinal static surface current modes are nontrivial solutions of the eigenvalue problem:
\begin{equation}
	\mathcal{T}_0^\| \{ \vec j^\|_k \} \left( {\vec r} \right)  =  \gamma_k^\|  \, \vec j^\|_k \qquad \text{on} \; \partial \Omega,
	\label{eq:AuxProb_Tl}
\end{equation}
where
\begin{equation}
		\mathcal{T}_0^\| \left\{ {\vec w} \right\} \left( {\vec r} \right) = \vers{n} \times \vers{n} \times \nabla \oint_{\partial\Omega} g_0 \left( {\vec r}  - {\vec r'} \right) \nabla' _S  \cdot {\bf w}\left( {\vec r}' \right) \dS',
	\label{eq:Operator_Tlo}
\end{equation}
$\nabla_{S} \cdot$ denotes the surface divergence, and $g_0$ is the homogeneous space static Green's function
\begin{equation}
	g_0 \left( {\vec r} - {\vec r}' \right) = \frac{1}{4 \pi} \frac{ 1}{ \|{\vec r} - {\vec r}' \| }.
	\label{eq:StaticGreen}
\end{equation}
 Apart from a multiplicative factor, the operator $\mathcal{T}_0^\|$ takes a surface electric current density \( \vec{w}\) defined on the boundary $\partial \Omega$ of the particle and outputs the tangential component of the electric field it generates on $\partial \Omega$ in the electrostatic limit. 
 
Eq. \eqref{eq:AuxProb_Tl} is numerically solved by using as expansion and test functions the \textit{star-basis} functions \cite{wilton_novel_1993, burton_study_1995,vecchi_loop-star_1999}.
Longitudinal static modes constitute a basis for the square-integrable functions defined on $\partial \Omega$ and having zero surface curl \cite{mayergoyz_plasmon_2013}, and diagonalize the first operator in Eq. \eqref{eq:EFIE_def} in the small-particle limit. These modes are the \textit{natural modes} of electromagnetic scattering from surfaces of finite conductivity  in the electrostatic limit \cite{forestiere_electromagnetic_2019}. 

The transverse static surface current modes are nontrivial solutions of the eigenvalue problem:
\begin{equation}
	\label{eq:AuxProb_Tt}
	\mathcal{T}_0^\perp \left\{ \vec j^\perp_k \right\} \left( {\vec r} \right)  = \gamma_k^\perp  \vec j^\perp_k \qquad \text{on} \; \partial \Omega,
\end{equation}
with 
\begin{equation}
	\label{eq:Operator_Tto}
		\mathcal{T}_0^\perp \left\{ {\vec w} \right\} \left( {\vec r} \right) =  -\vers{n} \times \vers{n} \times \oint_{\partial \Omega} g_0 \left( \vec r - \vec r' \right) {\vec w} \left( {\vec r}' \right) \dS'.
\end{equation} 
 Apart from a multiplicative factor, the operator $\mathcal{T}_0^\perp$ takes a surface electric current density \( \vec{w}\) defined on $\partial \Omega$ and outputs the tangential component of the electric field it generates on $\partial \Omega$ in the magnetostatic limit.
This equation is numerically solved by using as expansion and test function the \textit{loop functions} \cite{wilton_novel_1993, burton_study_1995, vecchi_loop-star_1999}.
Transverse static modes constitute a basis for the square integrable functions defined on $\partial \Omega$ having zero surface divergence (can be proved by following Ref. \cite{tamburrino_monotonicity_2021}), and diagonalize the second operator in the expression of $\mathcal{T}$ \eqref{eq:EFIE_def} in the small-particle limit. These modes are the \textit{natural modes} of electromagnetic scattering from surfaces of finite conductivity in the magnetostatic limit  \cite{forestiere_electromagnetic_2019} . 

Several characteristics make the use of static modes appealing. First, the retarded Green's function, serving as the kernel for integral operators in surface integral formulation, can be decomposed into the static Green's function, which includes the integrable singularity, and a distinct term that is a regular function. The integral operators that include the static Green's function are diagonalized by the static modes, thus regularizing the overall problem. Second, employing the expansion of static modes, along with suitable rescaling and rearrangement of the unknowns, renders the surface integral formulation resistant to the low-frequency breakdown issue. Third, static modes are solely dependent on the object's shape, allowing for their universal application across diverse operating frequencies and materials, under the assumption that the object is homogeneous. This enables a consistent, frequency-independent representation of any scattering scenario involving objects of a particular shape, surpassing other basis sets, such as characteristic modes \cite{bucci_use_1995,garbacz_modal_1965,chang_surface_1977,harrington_characteristic_1972,chen_characteristic_2015}, which vary with frequency and material properties.  Fourth,  for objects of size smaller than the wavelength of operation, only a few static modes are sufficient to accurately describe the emergent scattering response \cite{forestiere_static_2023}.

The transverse static modes $\{\vec{j}_n^\perp\}_{n=1}^{N^\perp}$, associated with the first $N^\perp$ eigenvalues $\gamma_n^\perp$ (sorted in descending order), and the longitudinal static modes $\{\vec{j}_m^\|\}_{m = 1}^{N^\|}$,
associated with the first $N^\|$ eigenvalues $\gamma_m^\|$ (sorted in ascending order) are used as entire domain basis function to represent the equivalent electric and magnetic surface currents on $\Omega_\partp$, namely $\mathcal{B}_\psi^\partp = \{\boldsymbol{\psi}^\partp_i\}_{i=1}^{N^\perp + N^\|} =  \{\vec{j}_1^\perp,...,\vec{j}_{N^{\perp}}^\perp,\vec{j}_1^\|,...,\vec{j}_{N^\|}^\|  \}$.

The matrices of Eq. \eqref{eq:PMCHWT_finite} are first evaluated in terms of the RWG representation. 
Then the transformation (\ref{eq:BasisTransformation}) can be used to pass from a RWG representation to the static modes representation. 

As the static surface modes are a very effective description of the surface current density field on objects whose dimensions are small compared to the wavelength, the dimension of the operators { $\{T\}_{\psi}$ and $\{K\}_\psi$ is typically smaller than $\{T\}_\xi$ and $\{K\}_\xi$, the transformation \eqref{eq:BasisTransformation} effectively acts as a \textit{compression}}. Consequently, the system \eqref{eq:PMCHWT_finite} now has $2(N^\perp + N^\|)\Npart$ degrees of freedom, which is typically drastically smaller than $2n_e\Npart$ derived from the RWG basis, where $n_e$ is the number of edges in the single-particle mesh. In the following, we denote as $\boldsymbol{\psi}$ the basis function in the static modes basis.
 
\section{Fast Multipole Method}

\label{sec:MLFMA}
We apply the Multilevel Fast Multipole Algorithm (MLFMA) \cite{greengard_fast_1987,engheta_fast_1992,coifman_fast_1993,song_multilevel_1997} to solve the PMCHWT system \eqref{eq:PMCHWT_finite} using the static mode basis to represent electric and magnetic surface currents. MLFMA accelerates the matrix-vector product by enabling groups of basis functions to interact collectively rather than through pairwise interactions between every possible pair of basis functions.

Initially, basis functions are grouped. Traditionally, grouping basis functions involves a hierarchical tree structure, such as a Quadtree or Octree \cite{chew_fast_2000, song_multilevel_1997}. The bounding box, i.e., the smallest box containing the entire object, is first identified. This box is then recursively subdivided until, at the finest level, the size of the box is a fraction of the wavelength, typically $\lambda/4$ \cite{chew_fast_2000}. 
\begin{figure}[t!]
	\centering
 \includegraphics[width = \linewidth]{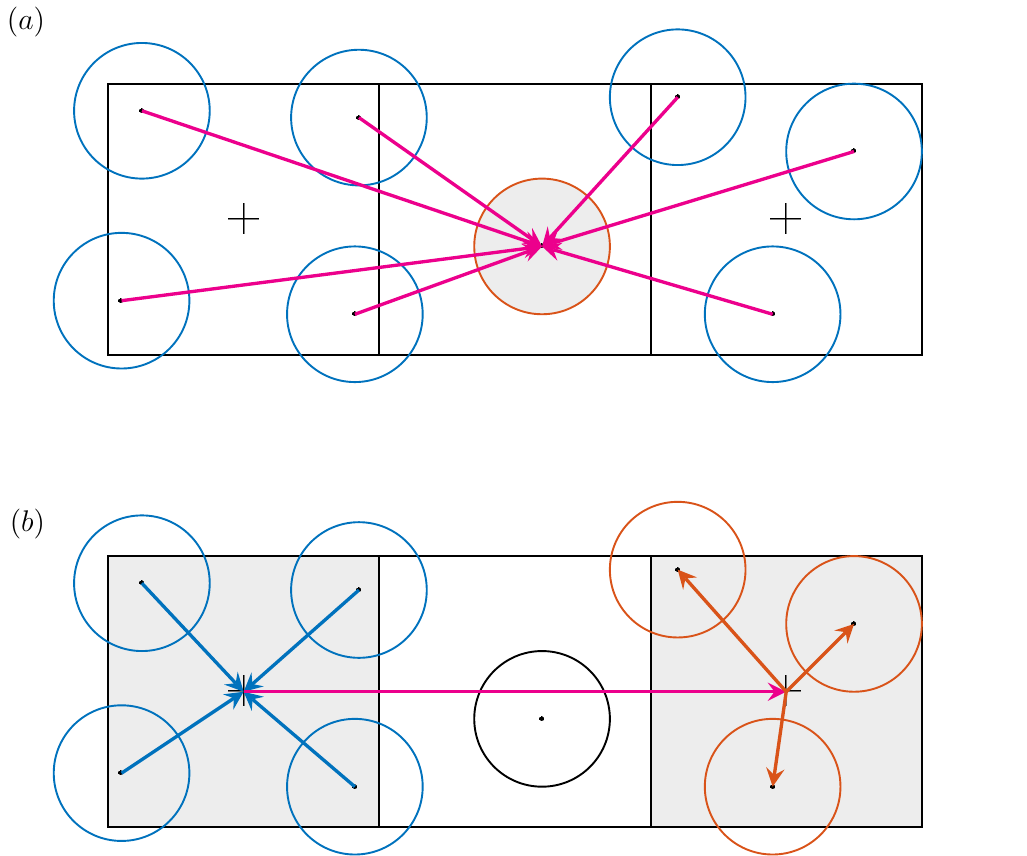}
\caption{Illustration of the far product in the MLFMA-SMR. (a) Particle-to-particle interaction on the finest level $L_{\max}$. (b) Level $L_{\max}-1$.}
 \label{fig:FMM_illustration}
\end{figure}
At the finest level, standard Quadtree/Octree approaches might divide a meta-atom into different groups, making the grouping of entire-domain basis functions ambiguous. Leveraging the fact that we deal with meta-atoms whose dimensions are smaller than the wavelength, we group basis functions by considering each particle as a finest-level group, as shown in Fig. \ref{fig:FMM_illustration}(a). At this level, all static modes $\boldsymbol{\psi}^\partp_i$ are physically associated with the center of the particle $\Omega_\partp$.  For coarser levels, we hierarchically group the basis functions using a standard Quadtree structure (Fig. \ref{fig:FMM_illustration}(b)), which is particularly suitable for structures where the lateral dimensions significantly exceed the vertical ones, as in the case of metasurfaces and metalenses.

Following this, the impedance matrix $Z$ is decomposed as the sum of two matrices: the near-field matrix, and the far-field matrix. The near-field matrix is associated with self-interaction and interactions among basis functions within adjacent groups, while the far-field matrix is associated with interactions between non-adjacent basis functions. When the particles are well separated, namely their distance is greater than $\lambda/2$, only the self-interactions are considered as near. Additionally, for identical particles, the self-particle blocks $Z^{\partp\partp}$ are identical, i.e. $Z^{\partp\partp} = Z_s$ for $\partp = 1,...,\Npart$. Thus, instead of evaluating and storing $Z^{\partp\partq}$ for all $\Omega_\partq$ near to $\Omega_\partp$, $Z_s$ can be computed and stored once, leading to significant savings in memory requirements and CPU time, compared to classical MLFMA implementation. To address the singularities in Green's function when $\vec{r}  = \vec{r}'$, the numerical integration of shape function times Green's function or its gradient are evaluated using the techniques introduced in Refs. \cite{graglia_numerical_1993} and \cite{hodges_evaluation_1997}.

According to \eqref{eq:Zpq_delta},  only the operators $T_-^{\partp\partq}$ and $K_-^{\partp\partq}$ appear in $Z^{\partp\partq}$ when  $\partp \neq \partq$. Thus, in the following we simplify the notations by omitting the ``$-$" subscript.
To efficiently evaluate the $(m,n)$ entry of the operators $T^{\partp\partq}$ and $K^{\partp\partq}$ when $\Omega_\partp$ is far from $\Omega_\partq$ (i.e. $\partp\neq \partq$), { the fast multipole method \cite{greengard_fast_1987} prescribes to rewrite the  matrices in Eq. \eqref{eq:PMCHWT_Finite1}  as}
\begin{equation}
T_{mn}^{\partp\partq}= -jk \int_{\partial \Omega_\partp} \boldsymbol{\psi}_m^\partp(\vec{r}) \cdot \int_{\partial\Omega_\partq} \dyad{G}(\vec r , \vec{r}')\boldsymbol{\psi}_n^\partq(\vec{r'})  \dS' \, \dS,
\label{T_far1}
\end{equation}
\begin{equation}
K_{mn}^{\partp\partq}=  \int_{\partial \Omega_\partp} \boldsymbol{\psi}_m^\partp(\vec{r}) \cdot  \int_{\partial\Omega_\partq}\boldsymbol{\psi}_n^\partq(\vec{r'})  \times \nabla ' g(\vec r - \vec{r}')   \dS' \, \dS
\label{K_far1},
\end{equation}
where $\dyad{G}$ is the dyadic Green's function
\begin{equation}
    \dyad{G}(\vec{r},\vec{r}')  = \left( \dyad{I} - \frac{1}{{k}^2} \nabla \nabla'\right) g(\vec{r}- \vec{r}'),
\end{equation}
and $\dyad{I}$ is the unit dyad. Then, using the addition theorem \cite{chew_fast_2000}, Green's function can be expressed as
\begin{equation}
\frac{e^{-jk\|\vec{r} - \vec{r}'\|}}{4\pi \|\vec{r}-\vec{r}'\|} \approx  \int_{\mathbb{S}^2} e^{-jk\vers{k}\cdot(\vec{r} - \vec r_\partp)} T_L(\vers{k}, \vec{r}_{\partp\partq})  e^{jk\vers{k}\cdot(\vec{r}' - \vec r_\partq)} \mathrm{d}\vers{k},
\label{eq:addition_th}
\end{equation}
with $\vec{r}\in \Omega_\partp$, $\vec{r}' \in \Omega_\partq$, $\vec{r}_{\partp\partq} = \vec{r}_\partp - \vec {r}_\partq$, where $\vec{r}_\partp$ and $\vec{r}_\partq$ are, respectively, the centers of the particles $\Omega_\partp$ and $\Omega_\partq$, $\mathbb{S}^2 = \{ \vers{k} \in \mathbb{R}^3 : \| \vers{k}\|_2 = 1\}$ is the unit sphere,  $T_L$ is the translation function \cite{chew_fast_2000, song_error_2001, hastriter_error_2003, song_multilevel_1997} truncated at the order $L$:
\begin{multline}
	T_L( \vers{k}, \vec{r}_{\partp\partq}) =  \\-\frac{jk}{(4\pi)^2} \sum_{\ell = 0}^{L} (-j)^{\ell} (2\ell +1) h_\ell^{(2)} (k\|\vec{r}_{\partp\partq}\|) P_\ell (\vers{k}\cdot \vers{r}_{\partp\partq}),
 \label{eq:translation_fun}
\end{multline}
where $h_\ell^{(2)}$ is the second kind spherical Hankel function of order $\ell$, $P_\ell$ is the Legendre Polynomial of order $\ell$. 
Using the expansion \eqref{eq:addition_th} in Eqs. \eqref{T_far1} and \eqref{K_far1}, we obtain

\begin{equation}
	T^{\partp\partq}_{mn} \approx  -jk\int_{\mathbb{S}^2} \boldsymbol{\hat{\psi}}^{\partp}_m(\vers{k})\cdot T_L(\vers{k}, \vec{r}_{\partp\partq})(\dyad{I} - \vers{k}\vers{k}) \cdot \boldsymbol{\hat{\psi}}^{\partq*}_n(\vers{k}) \, \mathrm{d}\vers{k},
 \label{eq:Tmn_fmm}
\end{equation}

\begin{equation}
	K^{\partp\partq}_{mn} \approx -jk\int_{\mathbb{S}^2} \vers{k} \times\boldsymbol{\hat{\psi}}^\partp_m(\vers{k})\cdot T_L(\vers{k}, \vec{r}_{\partp\partq})  \boldsymbol{\hat{\psi}}^{\partq*}_n(\vers{k})\ \mathrm{d}\vers{k},
  \label{eq:Kmn_fmm}
\end{equation}
where $\boldsymbol{\hat{\psi}}^\partp_m$ is the $k$-space representation of the basis function
\begin{equation}
\boldsymbol{\hat{\psi}}^\partp_m(\vers{k}) = \int_{\partial\Omega_\partp} \boldsymbol{\psi}_m^\partp(\vec{r}) e^{-jk\vers{k}\cdot (\vec r - \vec r_\partp)} \dS.
\label{eq:kspace_Rap}
\end{equation}
\begin{figure*}[t]
\centering
\includegraphics[width=1\textwidth]{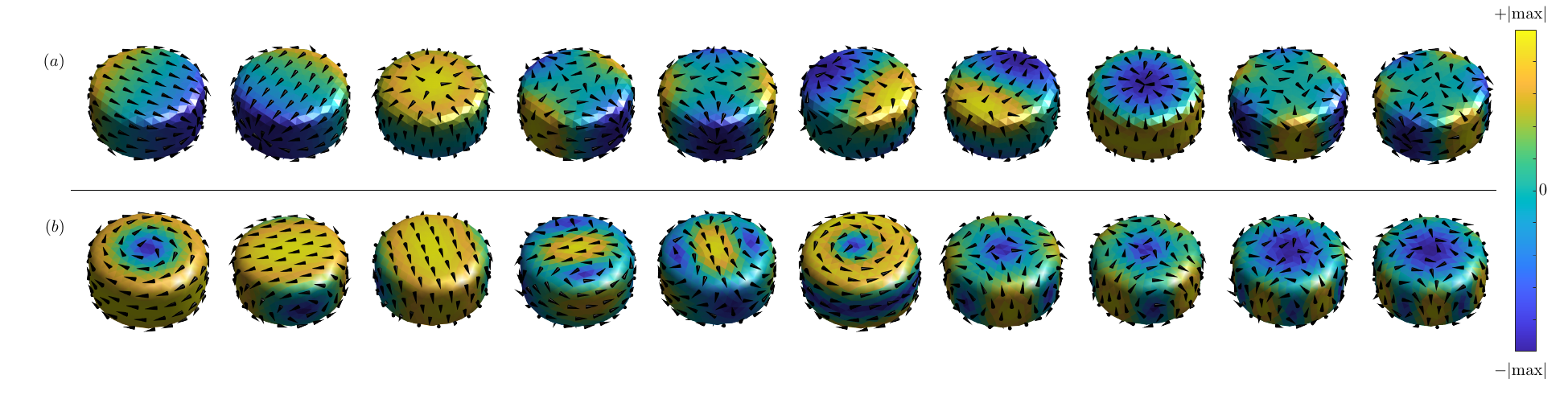}
\caption{(a) First ten longitudinal static modes of a nanodisk of radius 100 nm and height 50 nm, described by a surface mesh having 588 nodes, 1172 triangles, and 1758 edges.  The arrows represent the direction of the surface current density field, the color represents the surface charge density. (b) First ten transverse static modes of a nanodisk. The arrows represent the direction of the surface current density field, the color represents the magnitude of the current density field. Modes are presented in ascending lexicographic order and sorted according to their respective static eigenvalues. }
	\label{fig:TranLongModes}
 \end{figure*}
{In this context, $\boldsymbol{\hat{\psi}}_m^\partp$ and  $\boldsymbol{\hat{\psi}}_m^{\partp*}$ are also called \textit{receiving} and \textit{radiation patterns}, respectively.}

The truncation of the translation function \eqref{eq:translation_fun} at finite order $L$  depends on the electrical size of the groups, $ka$, where $a$ is the group size. An approximation for $L$ is given by \cite{hastriter_error_2003,chew_fast_2000} $L = kd + 1.8 d_0^\frac{2}{3}(kd)^{\frac{2}{3}}$, where $d_0$ is the desired number of significant digits, and $d = \sqrt{2}a$. However, truncating the series in  Eq. \eqref{eq:translation_fun} at a finite order $L$ presents challenges due to numerical instabilities caused by the nature of spherical Hankel functions \cite{abramowitz_handbook_1964,song_error_2001, hastriter_error_2003, dembart_accuracy_1998}. While increasing $L$ might intuitively seem to reduce error, the oscillatory nature of these functions at higher orders complicates this approach, often leading to diminished accuracy rather than improvement \cite{darve_efficient_2004}. Furthermore, these functions exhibit divergence at small arguments,  corresponding to cases where the electrical distance between two centers is very small, leading to a low-frequency breakdown \cite{song_error_2001,darve_efficient_2004}.  Nonetheless, in the context of the investigated metasurfaces, these instabilities are typically absent, as the particles under consideration are sufficiently separated. {
Furthermore, we can directly assess the maximum error in the Eqs. \eqref{eq:Tmn_fmm} and \eqref{eq:Kmn_fmm} as 
\begin{equation}
    \varepsilon_L\{A\} = \max_{{\partp,\partq}} \frac{\| A^{\partp \partq} - \bar{A}^{\partp \partq}\|_2}{\|\bar{A}^{\partp \partq}\|_2}
    \label{eq:error_TK}
\end{equation}
{ where $\| \cdot \|_2$ is the matrix 2-norm, $\bar{A}$ is the matrix $T$ or $K$ evaluated by the Eq. \eqref{eq:PMCHWT_Finite1}, while $A$ is the matrix $T$ or $K$ evaluated by the Eqs. \eqref{eq:Tmn_fmm}-\eqref{eq:Kmn_fmm}}.
}


\section{Results and Discussion}
\label{sec:Results}

The execution of the numerical algorithm can be subdivided into five stages: i) Assembly of the near-field impedance matrix $Z_N$, and the right-hand side vector $\vec{V}$ using RWG basis functions. ii) Generation of the static modes of an isolated object. Specifically, the longitudinal static modes are obtained by solving the eigenvalue problem (\ref{eq:AuxProb_Tl}) using star functions as both expansion and testing functions; the transverse static modes are obtained by solving the eigenvalue problem (\ref{eq:AuxProb_Tt}) using loop functions as both expansion and testing functions. As an example, in Fig. \ref{fig:TranLongModes}, we present the calculated first ten longitudinal and ten transverse static modes for a nanodisk of radius 100 nm and height 50 nm. iii)  Evaluation of the MLFMA operators, consisting of the translation function $T_L$, the receiving and radiation patterns \eqref{eq:kspace_Rap}. The time complexity and memory requirements for assembling $T_L$ are on the order of $\mathcal{O}(\Npart \log \Npart)$. {Since the radiation pattern \eqref{eq:kspace_Rap} is evaluated for each particle, his time and memory complexities are on the order of $\mathcal{O}(\Npart)$}. These complexities can be reduced to $\mathcal{O}(1)$ if the particles have the same spatial orientation (i.e., there is no rotation between particles). iv) Compression stage, i.e.transitioning from the representation of the near matrix and the radiation/receiving patterns in terms of the RWG basis to the representation in terms of the static modes basis. This stage can be integrated into steps i) and iii), as there is no need to directly store the RWG representation of these operators. v) Solution of the PMCHWT system using an iterative method accelerated by the MLFMA, which reduces the computational complexity of the matrix-vector product from $\mathcal{O}(\Npart^2)$ to $\mathcal{O}(\Npart \log \Npart)$. In particular, we used the Generalized Minimal Residual Method (GMRES) \cite{saad_gmres_1986}. This stage is the most time-consuming, as the entire system is solved with a complexity of $\mathcal{O}(I \Npart \log \Npart)$, where $I$ is the number of iterations required. To accelerate the convergence of the GMRES, we employ a block-diagonal preconditioner, where each block is the single-particle self-interaction impedance matrix $Z_s$. It effectively reduces $I$ to $\mathcal{O}(\Npart^\alpha)$, with $\alpha <1$. We choose a relative tolerance of $10^{-4}$.

\begin{figure*}[t]
\centering
\includegraphics[width=1\linewidth]{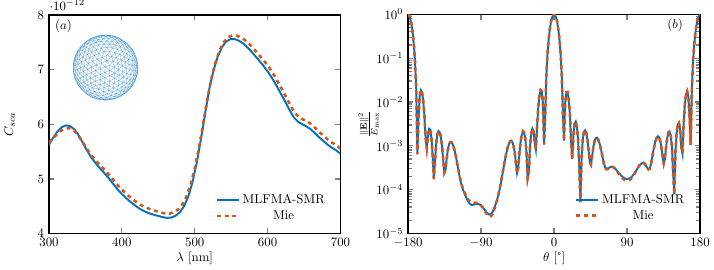}
 \caption{(a) Scattering cross-section of a finite Vogel spiral composed of 100 gold spheres of radius $R = 100$ nm under the excitation of a linearly polarized plane wave at wavelength $\lambda$. $C_{sca}$ is computed using the MLFMA-SMR with $N^\| = N^\perp = 10$. (b) Normalized squared magnitude of the scattered electric field in the far region at $\lambda$ = 600 nm. The reference Mie solution (dashed line) is also provided for comparison. (Inset) Surface mesh used to discretize the meta-atoms in the array, consisting of 500 nodes, 996 triangles, and 1494 edges.}
\label{fig:Csca_spectrum}
\end{figure*}

In the following simulations, the truncation number $L$ of the translation function \eqref{eq:translation_fun}, is initially set to $10$ at the finest level. This value is subsequently doubled at each successive level \cite{chew_fast_2000}. The numerical evaluation of the integrals over the unit sphere is performed using $K = 2L^2$ points \cite{chew_fast_2000}.
For interpolation, we employ a $4\times 4$ Lagrange interpolation scheme.
The FORTRAN code used to implement the entire algorithm operates on a single CPU (Intel Xeon-Gold 6140 M 2.3 GHz).

\subsection{Validation}

\label{sec:Validation}
We validate the MLFMA-SMR method by investigating the scattering from an  array consisting of 100 identical gold nano-spheres, each with a radius $R = 100$ nm. The particles are arranged according to a \textit{golden angle} spiral pattern \cite{trevino_circularly_2011}, where the center of the $i$-th particle is located at $(x_i,y_i) =  r_i (\cos(i  \alpha) , \sin(i  \alpha)) $, where $i  = 1,..., \Npart$; $r_i = c\sqrt{i}$, $c$ is a constant scaling factor, assumed $c = \sqrt{3}R$, $\alpha = 2\pi/\phi^2$ is the golden angle, and  $\phi = (1+\sqrt{5})/2$ is the golden ratio. This aperiodic pattern is particularly convenient for our study as it is deterministic, and allows us to increase the dimension of the array one particle at a time without appreciably changing the array's density.
The gold permittivity is obtained by interpolating experimental data \cite{johnson_optical_1972}. The array is illuminated by a linearly polarized plane, propagating along the axis orthogonal to the array's plane, linearly polarized along the $x$-axis. The wavelength $\lambda$ is varied within the interval $[300,700]$nm. Each particle's surface is discretized using the triangular mesh shown in the inset of Fig. \ref{fig:Csca_spectrum}(a). 

The scattering cross section $C_{sca}$ is plotted in Fig. \ref{fig:Csca_spectrum} as a function of $\lambda$. The solution, computed by the MLFMA-SMR, utilizing $N^\| = N^\perp = 10$ modes per particle (resulting in a total of $4000$ unknowns) is compared against the multiparticle Mie theory \cite{xu_electromagnetic_1995}. The field $\mathbf{E}_s^{(\nu)}$ scattered by the $\nu$-th particle is described in terms of transverse magnetic (TM)  $\mathbf{N}_{m n}^{(3)}$, and transverse electric (TE) $\mathbf{M}_{m n}^{(3)}$ vector spherical wave functions \cite{bohren_absorption_2008}, which are regular at infinity and have pole at the $\nu$-th particle's center:

$$
\mathbf{E}_s^{(\nu)}(\mathbf{r})=\sum_{n=1}^{N_{\max }} \sum_{m=-n}^n a_{m n}^{(\nu)} \, \mathbf{M}_{m n}^{(3)}(\mathbf{r})+b_{m n}^{(\nu)} \, \mathbf{N}_{m n}^{(3)}(\mathbf{r}),
$$ 
where $a_{m n}^{(\nu)}$ and $b_{m n}^{(\nu)}$ are the expansion coefficient. We assume $N_{\max} = 15$.

The relative error on the scattering cross section is defined as $ \epsilon\{C_{sca}\} = {|C_{sca} - \bar{C}_{sca}|}/{\bar{C}_{sca}}$, where $\bar{C}_{sca}$  is the reference solution obtained by the multiparticle Mie theory. 
The achieved error is lower than $0.02$ all over the investigated spectral range. For the same array, in Fig. \ref{fig:Csca_spectrum}(b) we show the normalized squared magnitude of the (far) scattered electric field at a given wavelength $\lambda = 600$ nm. The reference Mie solution, obtained for $N_{\max} = 30$, is included for comparison.  An excellent agreement is found.

\subsection{Computational time and memory cost}
\begin{figure}[t]
	\centering
\includegraphics[width=1\linewidth]{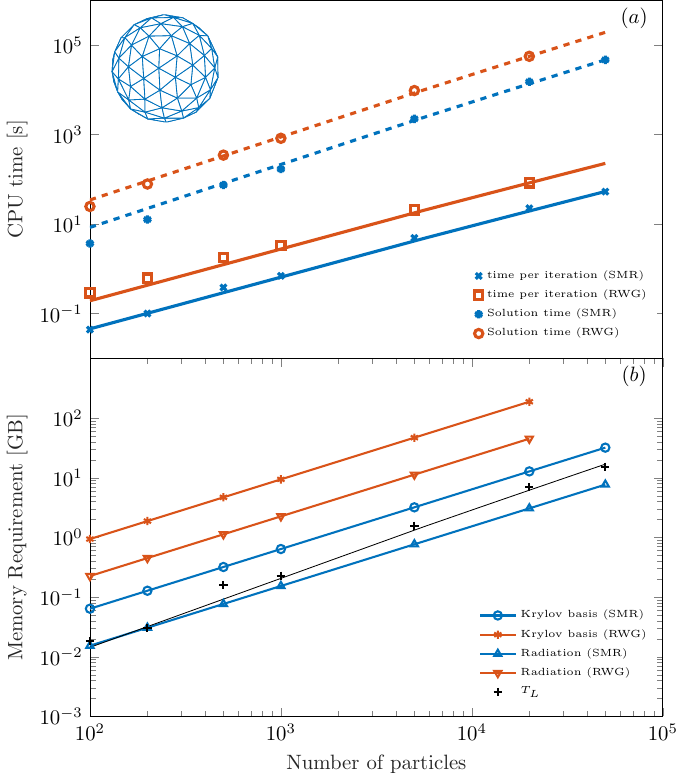}
 \caption{(a) CPU time required for the MLFMA solution of the PMCHWT system using RWG with static mode representation. The code runs on a single CPU. (b) Memory requirement comparison between the RWG with static mode representation. The $\mathcal{O}(\Npart \log \Npart)$ (continuous line) and $\mathcal{O}\left( \Npart^{1.25} \log \Npart\right)$ (dashed line) curves are also plotted for reference. (Inset) Surface mesh used for the computational complexity test.}
 	\label{fig:scaling_time_mem}
\end{figure}
We conducted a complexity study on metasurfaces consisting of a varying number up to $5\cdot 10^4$ of identical gold nano-spheres, each of radius R = 100 nm. The particles are arranged according to the golden-angle spiral pattern, defined in Sec. \ref{sec:Validation}.   The array is excited by a linearly polarized plane wave of wavelength $\lambda = 600$ nm, propagating along the direction orthogonal to the array's plane. Each sphere has been discretized by the surface mesh shown in the inset of Fig. \ref{fig:scaling_time_mem} (a), with $100$ nodes, $196$ triangles, $294$ edges, which results in an average edge length approximately equal to $ \lambda/15$.

In Fig. \ref{fig:scaling_time_mem}, we compare the CPU time and memory requirements for simulations using the multilevel fast multipole algorithm with RWG basis functions, denoted as MLFMA-RWG, and with the static modes basis, denoted as MLFMA-SMR, with $N^| = N^\perp = 10$. The results demonstrate that, in both cases of RWG and SMR, the MLFMA reduces the computational complexity of the matrix-vector product from $\mathcal{O}(\Npart^2)$ to $\mathcal{O}(\Npart \log \Npart)$. The degrees of freedom and the computational time for each case are detailed in Tab. \ref{tab:table_comparison}. Notably, the SMR-based computation is approximately 3 to 5 times faster than the corresponding computation using RWG basis functions. It is important to highlight that, for both RWG and SMR, the error compared to Mie theory is comparable, with an approximate error $\varepsilon \left\{ C_{sca} \right\} \approx 3\%$.

However, the speed-up depends on various factors, including the number of points used for interpolation/anterpolation between different levels of the quadtree, or the number of points used to sample the transfer function $T_L$ on the unit sphere. Above all, mesh density on each particle plays a very important role. Indeed, the advantages introduced by the MLFMA-SMR become much more apparent with denser meshes, necessary for discretizing irregular objects, such as nanofins and nanorods typically used in metalens.

In the investigated example, the speed-up is less than the number of unknowns' compression ratio $\kappa = n_e/(N^\parallel + N^\perp)$ between the number of basis function used in the RWG and the SMR cases. This fact is attributable to the computational time required by $k$-space representation of the MLFMA operators, which means that the computational complexity is not only linked to the number of unknowns but also to the electrical size of the array.

\begin{table}[h]
		\centering
		\caption{Comparison between RWG and SMR.}
  
\scriptsize 
\begin{tabular}{|c|c|c|c|c|c|c|c|c|}
			\hline
			 & \multicolumn{2}{c|}{{Total Number}} & \multicolumn{2}{c|}{{Time per}} & \multicolumn{2}{c|}{{Total GMRES}} & \multicolumn{2}{c|}{{Number of}} \\
		$p$	& \multicolumn{2}{c|}{{of Unknowns}} & \multicolumn{2}{c|}{{iteration [s]}} & \multicolumn{2}{c|}{{time [min]}} & \multicolumn{2}{c|}{{iterations}} \\
			\cline{2-9}
			& {RWG} & {SME} & {RWG} & {SME} & {RWG} & {SME} & {RWG} & {SME} \\
			\hline
			100 & $5.880 \cdot 10^4$ & $4\cdot 10^3$ & 0.29 & 0.04 & 0.42 & 0.06 & 85 & 84 \\
			\hline
			200 & $1.176\cdot 10^5$ & $8\cdot 10^3$ & 0.62 & 0.10 & 1.32 & 0.21 & 127 & 126 \\
			\hline
			500 & $2.940 \cdot 10^5$ & $2\cdot 10^4$ & 1.78 & 0.38 & 5.82 & 1.25 & 196 & 195 \\
			\hline
			1000 & $5.880 \cdot 10^5$ & $4\cdot 10^4$ & 3.37 & 0.70 & 13.84 & 2.85 & 246 & 243 \\
			\hline
			5000 & $2.940\cdot 10^6$ & $2\cdot 10^5$ & 20.95 & 4.92 & 161.34 & 37.61 & 462 & 458 \\
			\hline
			20000 & $1.176 \cdot 10^7$ & $8\cdot 10^5$ & 83.88 & 22.67 & 939.46 & 253.51 & 672 & 671 \\
			\hline
			50000 & - & $2\cdot 10^6$ & - & 53.32 & - & 782.08 & - & 880 \\
			\hline
		\end{tabular}
		\label{tab:table_comparison}
\end{table}
\begin{figure}[ht!]
    \centering
    \includegraphics[width=1\linewidth]{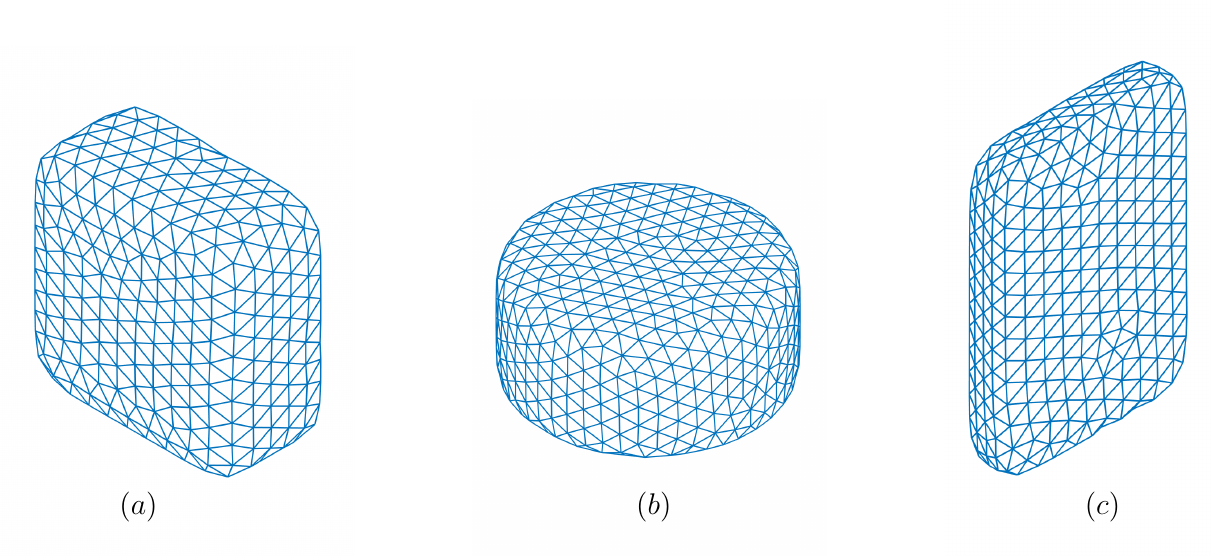}
    \caption{(a) Surface mesh of a brick, consisting of 436 nodes, 868 triangles, and 1302 edges. (b) Surface mesh of a nanodisk, consisting of 588 nodes, 1172 triangles, and 1758 edges. (c) Surface mesh of a nanofins, consisting of 468 nodes, 932 triangles, and 1398 edges.}
    \label{fig:matrix_mesh}
\end{figure}

Another key advantage of using the SMR in the MLFMA is the significant reduction in memory requirements. Although the translation function $T_L$ is the same in both RWG and SMR, as it doesn't depend on the basis function, the matrices representing the radiation and receiving patterns are reduced in size by a factor of $\kappa$, the self-impedance matrix $Z_s$ is reduced by $\kappa^2$, where $\kappa$ is the compression ratio.

\begin{figure*}[t]
\centering
	\includegraphics[width=1\textwidth]{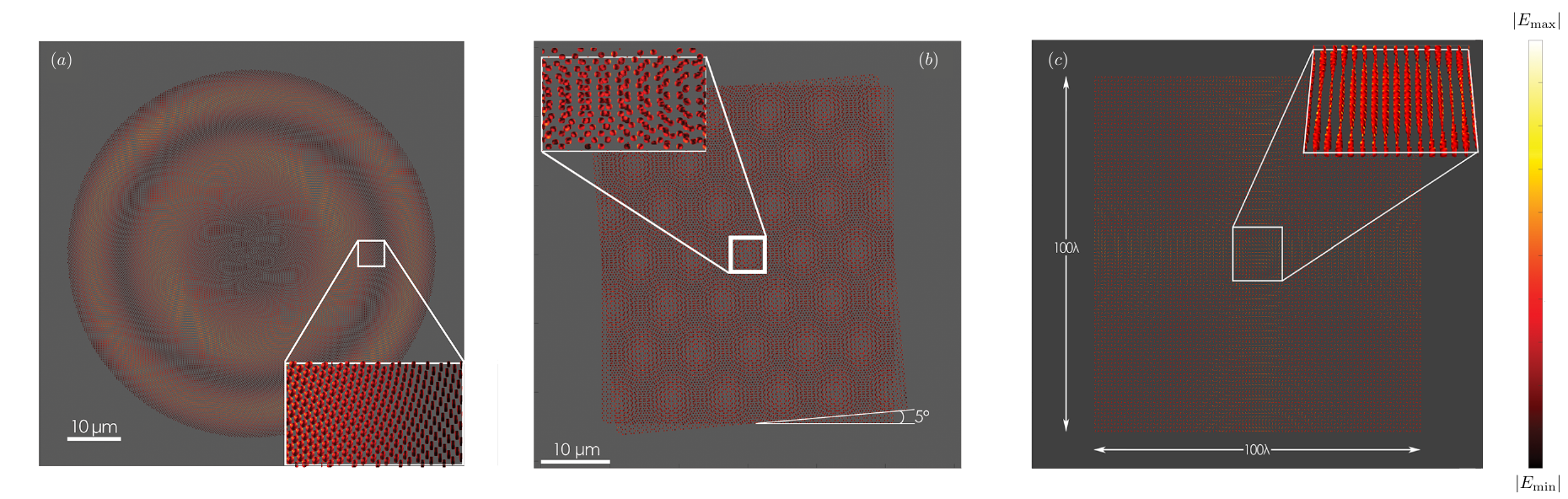}
 \caption{
 (a) Electric field magnitude on the particles' surface of a finite Vogel spiral composed of $40\cdot 10^{3}$ gold bricks of dimension $200 \times 200 \times 100$ nm under the excitation of a linearly polarized plane wave at wavelength $\lambda$ = 600 nm. The array size is roughly $115\lambda$. (b) Electric field magnitude on the particles' surface of a Moiré superlattice array composed of 20160 gold nanodisks of radius $100$ nm under the excitation of a linearly polarized plane wave at wavelength $\lambda$ = 600 nm. The array size is roughly $(76\times 87)\lambda$.  (c) Electric field magnitude on the particles' surface of a canonical metalens, lying on the xy plane, and composed of $10^4$ $\mathrm{TiO}_2$ nanofins of dimension $410\times 85\times 600$ nm under the excitation of a circularly polarized plane wave at wavelength $\lambda = 660$ nm. The array size is   $100\lambda$. In all three cases investigated, the solution has been computed by using the MLFMA-SMR with $N^\| = N^\perp = 10$.}	\label{fig:MatrixSurfaceFields}
\end{figure*}

Additionally, the memory needed to compute the GMRES solution after \(m\) iterations is proportional to the size of the Krylov subspace \(\mathcal{K}_m\), leading to a memory requirement of \(\mathcal{O}(mN)\), where \(N\) is the total number of unknowns. As \(N\) increases, the storage cost can become prohibitive, even for a relatively small number of iterations \(m\). The SMR method significantly reduces \(N\), enabling the simulation of larger structures that were previously infeasible due to memory constraints.

\subsection{Electrically large metasurfaces}

The use of the MLFMA-SMR finds its natural application in the numerical solution of the electromagnetic scattering problem by large metasurfaces, where an array, whose dimension can be much larger than the incident wavelength, is made of $\Npart$ objects with identical shape but with different orientation and size. We show three exemplificative case studies: a golden angle spiral of gold bricks, and a Moirè lattice of gold disks, and a canonical metalens of $\text{TiO}_2$ nano-fins.

In all three cases, we model the shape of the meta-atoms as a superellipsoid, whose boundary has the implicit equation 
\begin{equation}
    \left[ \left(\frac{x}{a_x}\right)^{\frac{2}{m}} + \left(\frac{y}{a_y}\right)^{\frac{2}{m}}\right]^{\frac{m}{n}} + \left(\frac{z}{a_z}\right)^{\frac{2}{n}} - 1 = 0,
    \label{eq:SuperEllipsoid}
\end{equation}
We used the public domain code developed by Per-Olof Persson and Gilbert Strang \cite{persson_simple_2004} to generate the surface mesh. 

We first consider a finite-size golden angle spiral array, { as defined in Sec. \ref{sec:Validation}}, having linear dimension of $115 \lambda$. The array is made of $40 \cdot 10^3$ gold bricks of dimension $200 \times 200 \times 100$ nm. We model this shape as a superellipsoid, using Eq. \eqref{eq:SuperEllipsoid} with $m = n = 0.25$, $a_x = a/2$, $a_y = a_z = a$. Each particle is described by the surface mesh shown in Fig. \ref{fig:matrix_mesh} (a). The $i$-th particle is rotated with respect to $\vers{x}$ by an angle $\theta_i = 2\pi (i-1)/\Npart$. The array is excited by a plane wave, linearly polarized in the plane of the array, along the $\hat{\mathbf{x}}$-axis, and propagating in the direction orthogonal to the array's plane with wavelength $\lambda = 600$ nm. The array size is about $70 \mu$m. In Fig. \ref{fig:MatrixSurfaceFields}(a), we show the electric field magnitude on the particles' surface. The solution is computed by using the MLFMA-SMR using the static mode representation with $N^\| = N^\perp = 10$, which corresponds to 40 unknowns per particle. The total computational time is about 20.4 hours on a single CPU. {The GMRES required approximately 15 hours and converged after 921 iterations.}

Next, we now consider an example of 2.5D metastructures, which are stacked layers of interacting objects. They provide sufficient degrees of freedom to implement efficient multifunctional devices. Specifically, we consider a two-layer Moiré superlattice made by a total number of $20160$ gold nanodisks of radius $R = 100$ nm and height $50$ nm. We model the nanodisk as a superellipsoid, whose boundary has the implicit equation \eqref{eq:SuperEllipsoid}, with $m = 1$, $n = 0.25$, $a_x = a_y = R$, $a_z = R/2$.  Each particle is described by the surface mesh shown in Fig. \ref{fig:matrix_mesh} (b). The centers of the particles are arranged at the vertices of two layers, obtained by periodically replicating a hexagon with a side length of $400$ nm. The second layer is located at a height $h = 300$ nm above the first and is rotated by an angle $\theta = 5^{\circ}$ with respect to it. The array is excited by a plane wave, linearly polarized in the plane of the array, along the $\vers{x}$-axis, and propagating in the direction orthogonal to the array's plane with wavelength $\lambda = 600$ nm. The array size is about $46 \times 53 \, \mu$m, which is roughly $(76\times 87)\lambda$. The solution is computed using the MLFMA-SMR with $N^\| = N^\perp = 10$. The total computational time is less than 6 hours on a single CPU.  {The GMRES required approximately 5.2 hours and converged after 564 iterations.}

\begin{figure}
    \centering
    \includegraphics[width=\linewidth]{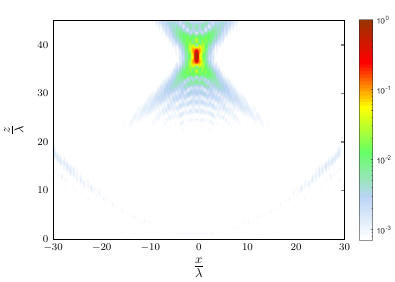}
    \caption{Electric field normalized intensity on the $x-z$ cross-sectional plane.  The achieved focal length is consistent with the prescribed value of $f=37.5\lambda$.  }
    \label{fig:focal_xz}
\end{figure}

Eventually, we consider an example of a canonical metalens \cite{hughes_perspective_2021}, composed of $10^4$ $\mathrm{TiO}_2$ nanofins of dimension $410\times 85\times 600$ nm. We model this shape as a superellipsoid, whose boundary has the implicit equation \eqref{eq:SuperEllipsoid}, with $m = n = 0.25$. Each particle is described by the surface mesh shown in Fig. \ref{fig:matrix_mesh} (c).  To achieve the desired phase profile, the $i$-th particle, centered at $(x_i,y_i)$, is rotated with respect to $\vers{x}$
by an angle \cite{hughes_perspective_2021, khorasaninejad_metalenses_2016}
\begin{equation}\theta_i = \frac{\pi}{\lambda}\left(f - \sqrt{f^2 + x_i^2 + y_i^2}\right),
\end{equation} where $f$ is the focal length. The metalens has a desired numerical aperture (NA) of 0.8, resulting in a focal length $f = 37.5\lambda$. The array is excited by a plane wave, circularly polarized in the plane of the array and propagating in the direction orthogonal to the array's plane with wavelength $\lambda = 660$ nm. The array size is $66 \times 66 \, \mu$m, which is  $(100\times 100)\lambda$. The solution is computed using the MLFMA-SMR with $N^\| = N^\perp = 10$.  The total computational time is about 16.8 hours on a single CPU. { The GMRES required approximately 10.6 hours and converged after 1019 iterations}.  In Fig. \ref{fig:MatrixSurfaceFields}(c) we show the electric field magnitude on the particles' surface.  In Fig. \ref{fig:focal_xz}(d) we show the normalized intensity $\|\vec{E}(x,z)\|^2$ on the $x$-$y$ cross-sectional plane. The achieved focal length is consistent with the prescribed value. 


\section{Conclusion}
In this work, we proposed a fast integral equation solver suitable for large particle arrays, even hundreds of wavelengths in diameter, comprising arbitrarily shaped objects with varying orientations and sizes. These scenarios are commonly encountered in the electromagnetic modeling of metasurfaces and metalenses. Our approach combines the Multilevel Fast Multipole Algorithm (MLFMA) with the Poggio-Miller-Chang-Harrington-Wu-Tsai (PMCHWT) formulation, utilizing a particular class of entire domain basis functions defined on the isolated particle, termed static modes \cite{forestiere_static_2023}. The static mode representation (SMR) significantly reduces the number of unknowns compared to discretizations based on sub-domain basis functions, such as RWG, thereby lowering both CPU time and memory requirements.

We validated the accuracy of the proposed MLFMA-SMR method by analyzing the scattering from finite aperiodic arrays of nano-spheres and comparing the results with those obtained using the multiparticle Mie theory \cite{xu_electromagnetic_1995}. Then, we report the CPU-time and memory requirements as functions of the number of particles composing the array, comparing the static mode representation to a more traditional approach using RWG basis functions. Our findings indicate that the MLFMA-SMR is $3$ to $5$ times faster and requires about $5$ times less memory than the traditional MLFMA employing RWG basis functions, while maintaining comparable accuracy. Furthermore, these speed and memory advantages significantly improve as the density of the surface mesh increases.

Eventually, we showcase the capabilities of the MLFMA-SMR method on different arrays: a golden-angle spiral array with a diameter of $115 \lambda$, consisting of 40,000 gold bricks; a two-layer Moiré superlattice made up of 20,160 gold nanodisks, with overall dimensions of $76\lambda \times 87\lambda$; and a canonical metalens composed of 10,000 $\mathrm{TiO}_2$ nanofins, covering an area of $100\lambda \times 100\lambda$ and having a focal length $f=37.5\lambda$.

Our results demonstrate that the MLFMA-SMR significantly reduces the total number of unknowns, leading to substantial savings in CPU time and memory requirements for the numerical solution of scattering problems from arrays of identical particles, compared to the classical MLFMA implementation with sub-domain basis functions. This advancement makes it feasible to simulate and design large metasurfaces.

\begin{acknowledgements}
This work was supported by the Italian Ministry of University and Research under the PRIN-2022, Grant Number 2022Y53F3X ``Inverse Design of High-Performance Large-Scale Metalenses".
\end{acknowledgements}

\bibliographystyle{ieeetr}

\end{document}